

\input harvmac

\input amssym.def
\input amssym
\baselineskip 14pt
\magnification\magstep1
\parskip 6pt

\input epsf

\font \bigbf=cmbx10 scaled \magstep1

\newdimen\itemindent \itemindent=32pt
\def\textindent#1{\parindent=\itemindent\let\par=\resetpar%
\indent\llap{#1\enspace}\ignorespaces}

\let\oldpar=\par
\def\resetpar{\oldpar\parindent=20pt\let\par=\oldpar}

\font\ninerm=cmr9 \font\ninesy=cmsy9
\font\eightrm=cmr8 \font\sixrm=cmr6
\font\eighti=cmmi8 \font\sixi=cmmi6
\font\eightsy=cmsy8 \font\sixsy=cmsy6
\font\eightbf=cmbx8 \font\sixbf=cmbx6
\font\eightit=cmti8
\def\eightpoint{\def\rm{\fam0\eightrm}
  \textfont0=\eightrm \scriptfont0=\sixrm \scriptscriptfont0=\fiverm
  \textfont1=\eighti  \scriptfont1=\sixi  \scriptscriptfont1=\fivei
  \textfont2=\eightsy \scriptfont2=\sixsy \scriptscriptfont2=\fivesy
  \textfont3=\tenex   \scriptfont3=\tenex \scriptscriptfont3=\tenex
  \textfont\itfam=\eightit  \def\it{\fam\itfam\eightit}%
  \textfont\bffam=\eightbf  \scriptfont\bffam=\sixbf
  \scriptscriptfont\bffam=\fivebf  \def\bf{\fam\bffam\eightbf}%
  \normalbaselineskip=9pt
  \setbox\strutbox=\hbox{\vrule height7pt depth2pt width0pt}%
  \let\big=\eightbig  \normalbaselines\rm}
\catcode`@=11 %
\def\eightbig#1{{\hbox{$\textfont0=\ninerm\textfont2=\ninesy
  \left#1\vbox to6.5pt{}\right.\n@@space$}}}
\def\vfootnote#1{\insert\footins\bgroup\eightpoint
  \interlinepenalty=\interfootnotelinepenalty
  \splittopskip=\ht\strutbox %
  \splitmaxdepth=\dp\strutbox %
  \leftskip=0pt \rightskip=0pt \spaceskip=0pt \xspaceskip=0pt
  \textindent{#1}\footstrut\futurelet\next\fo@t}
\catcode`@=12 %

\def\a{\alpha}
\def\b{\beta}
\def\c{\gamma}
\def\d{\delta}
\def\e{\epsilon}

\def\l{\lambda}
\def\m{\mu}
\def\n{\nu}

\def\r{\rho}
\def\s{\sigma}

\def\w{\omega}

\def\C{\Gamma}
\def\D{\Delta}

\def\W{\Omega}

\def\pl{\partial}
\def\ul{\underline}
\def\rta{\rightarrow}

\def\vp{v_{\rm ph}}
\def\vf{v_{\rm wf}}
\def\or{\overrightarrow}
\def\Dslash{\,{\raise.15ex\hbox{/}\mkern-12mu D}}

\lref\DH{I.T. Drummond and S. Hathrell, Phys. Rev. D22 (1980) 343. }
\lref\Drum{I.T. Drummond, Phys. Rev. D63 (2001) 043503. } 
\lref\Sone{R.D. Daniels and G.M. Shore, Nucl. Phys. B425 (1994) 634. }
\lref\Stwo{R.D. Daniels and G.M. Shore, Phys. Lett. B367 (1996) 75. }
\lref\Sthree{G.M. Shore, Nucl. Phys. B460 (1996) 379. }
\lref\Sfour{G.M. Shore, Nucl. Phys. B605 (2001) 455. }
\lref\Sfive{G.M. Shore, in preparation.}
\lref\Gibb{G.W. Gibbons and C.A.R. Herdeiro, Phys. Rev. D63 (2001) 064006.}
\lref\LPT{J. I. Latorre, P. Pascual and R. Tarrach, Nucl. Phys. B437
(1995) 60.}
\lref\Gies{W. Dittrich and H. Gies, Phys. Lett. B431 (1998) 420-429;
Phys. Rev. D58 (1998) 025004.}
\lref\Myers{R. Lafrance and  R.C. Myers, Phys. Rev. D51 (1995) 2584.}
\lref\Cho{H.T. Cho, Phys. Rev. D56 (1997) 6416.}
\lref\Cai{R-G. Cai, Nucl. Phys. B524 (1998) 639.}
\lref\Bass{B.A. Bassett, S. Liberati and C. Molina-Paris, Phys. Rev. D62 (2000) 103518.}
\lref\Scharn{K. Scharnhorst, Phys. Lett. B236 (1990) 354.}
\lref\Bart{G. Barton, Phys. Lett. B237 (1990) 559.}
\lref\LSV{S. Liberati, S. Sonego and M. Visser, gr-qc/0107091.}
\lref\LSVtwo{S. Liberati, S. Sonego and M. Visser, Phys. Rev. D63 (2001) 085003.}
\lref\BLV{M. Visser, B. Bassett and S. Liberati, Nucl. Phys. Proc. Suppl. 88
(2000) 267.}
\lref\DN{A.D. Dolgov and I.D. Novikov, Phys. Lett. B442 (1998) 82.}
\lref\DNtwo{A.D. Dolgov and I.D. Novikov, Phys. Lett. A243 (1998) 117.}
\lref\DK{A.D. Dolgov and I.B. Khriplovich, Sov. Phys. JETP 58(4) (1983) 671.}
\lref\Khrip{I.B. Khriplovich, Phys. Lett. B346 (1995) 251.}         
\lref\Konst{M.Yu. Konstantinov, gr-qc/9810019.}
\lref\Hau{L.V. Hau, Scientific American, July 2001, 66.}
\lref\WKD{L.J. Wang, A. Kuzmich and A. Dogoriu, Nature 406 (2000)  277.}
\lref\Brill{L. Brillouin, {\it Wave Propagation and Group Velocity}, 
Academic Press (London) 1960.}
\lref\Leon{M.A. Leontovich, {\it in} L.I. Mandelshtam,
{\it Lectures in Optics, Relativity and Quantum Mechanics}, Nauka, Moscow 1972~
{(\it in Russian).}}
\lref\HE{S.W. Hawking and G.F.R. Ellis, {\it The Large Scale Structure of
Spacetime}, Cambridge University Press, 1973.}
\lref\Fried{F.G. Friedlander, {\it The Wave Equation on a Curved Spacetime},
Cambridge University Press, 1975.}
\lref\CH{R. Courant and D. Hilbert, {\it Methods of Mathematical Physics, Vol II},
Interscience, New York, 1962.}
\lref\SEF{P. Schneider, J. Ehlers and E.E. Falco, {\it Gravitational Lenses},
Springer-Verlag, New York, 1992.}
\lref\Ch{S. Chandresekhar, {\it The Mathematical Theory of Black Holes},
Clarendon, Oxford, 1985. }
\lref\Inverno{R.A. d'Inverno, {\it Introducing Einstein's Relativity},
Clarendon, Oxford, 1992. }
\lref\Bondi{H. Bondi, M.G.J. van der Burg and A.W.K. Metzner, Proc. Roy. Soc.
A269 (1962) 21.}
\lref\Sachs{R.K. Sachs, Proc. Roy. Soc. A270 (1962) 103. }
\lref\TEone{W. Tsai and T. Erber, Phys. Rev. D10 (1974) 492.}
\lref\TEtwo{W. Tsai and T. Erber, Phys. Rev. D12 (1975) 1132.}
\lref\Adler{S. Adler, Ann. Phys. (N.Y.) 67 (1971) 599.}
\lref\BGZV{A.O. Barvinsky, Yu.V. Gusev, V.V. Zhytnikov and G.A. Vilkovisky,
Print-93-0274 (Manitoba), 1993.}
\lref\BV{A.O. Barvinsky and G.A. Vilkovisky, Nucl. Phys. B282 (1987) 163;
B333 (1990) 471; B333 (1990) 512.}


{\nopagenumbers
\rightline{SWAT/333}
\vskip1cm
\centerline{\bigbf Faster than Light Photons in Gravitational Fields II}
\vskip0.3cm
\centerline{\bigbf -- Dispersion and Vacuum Polarisation}

\vskip1cm

\centerline {\bf G.M. Shore}

\vskip0.5cm
\centerline{\it Department of Physics}
\centerline{\it University of Wales, Swansea}
\centerline{\it Singleton Park}
\centerline{\it Swansea, SA2 8PP, U.K.}

\vskip1cm

{
\parindent 1.5cm{

{\narrower\smallskip\parindent 0pt
ABSTRACT:~~Vacuum polarisation in QED in a background gravitational field
induces interactions which effectively violate the strong equivalence 
principle and affect the propagation of light. In the low frequency limit,
Drummond and Hathrell have shown that this mechanism leads to superluminal
photon velocities. To confront this phenomenon with causality, however, it
is necessary to extend the calculation of the phase velocity $\vp(\w)$ to high
frequencies, since it is $\vp(\infty)$ which determines the characteristics
of the effective wave equation and thus the causal structure. In this paper,
we use a recently constructed expression, valid to all orders in a derivative 
expansion, for the effective action of QED in curved spacetime to determine
the frequency dependence of the phase velocity and investigate whether 
superluminal velocities indeed persist in the high frequency limit.

\narrower}}}

\vskip2cm

\leftline{SWAT/333} 
\leftline{March 2002}

\vfill\eject}

\pageno=1

\newsec{Introduction}

It has been known since the original work of Drummond and Hathrell \refs{\DH} 
that quantum effects have important consequences for the propagation of light
in curved spacetime. In the classical theory of electrodynamics coupled
to general relativity, light propagates simply along null geodesics.
In quantum electrodynamics, however, vacuum polarisation changes the picture
and the background gravitational field becomes a dispersive medium for
the propagation of photons. In itself, this is perhaps not surprising.
The one-loop vacuum polarisation contribution to the photon propagator
introduces a non-trivial length scale $\l_c$ (the inverse electron
mass) and it is natural that photon propagation will be significantly
affected when the typical curvature scale $L$ is comparable to $\l_c$.
However, the remarkable result found by Drummond and Hathrell is
that in many cases the effect of vacuum polarisation is to induce
a change in the velocity of light to `superluminal' speeds, i.e.
$v > c$, where $c$ is the fundamental constant.

Since this original discovery, many special cases have been studied
in detail \refs{\DH,\Sone,\Stwo,\Sfour},
including propagation in black hole spacetimes described
by the Schwarzschild, Reissner-Nordstr\"om or Kerr metrics,
the FRW metric of big bang cosmology, and gravitational wave backgrounds,
in particular the Bondi-Sachs metric describing asymptotic radiation from
an isolated source\foot{Other examples and a selection of related work 
may be found in refs.\refs{\Myers,\Cho,\Cai,\DNtwo,\Bass,\Drum} and possible 
implications for time machines in \refs{\DN,\Konst,\LSV}.}.
The phenomenon of superluminal propagation has been 
observed in all these examples of gravitational fields and a number of
general features have been identified. A notable result is the 
`horizon' \refs{\Sthree} or `touching' \refs{\Gibb} theorem, 
which shows that even in the
presence of superluminal velocities, the effective black hole event horizon
defined by physical photon propagation coincides precisely with the
geometric horizon. Another important observation is that
the speed of light increases rapidly (as $1/t^2$ in the radiation dominated 
era) in the early stages of a FRW cosmology \refs{\DH}, 
with potential implications for the horizon problem and related issues. 

All this work has been based on the initial Drummond-Hathrell analysis,
in which they showed that the effect of one-loop vacuum polarisation
is to induce the following effective action, generalising the free-field
Maxwell theory:
\eqnn\sectaa
$$\eqalignno{
\C = &\int dx \sqrt{-g}\biggl[
-{1\over4}F_{\m\n}F^{\m\n} \cr
&~~~~+ {1\over m^2}\biggl(
a R F_{\m\n}F^{\m\n} + b R_{\m\n} F^{\m\l} F^\n{}_\l
+ c R_{\m\n\l\r} F^{\m\n} F^{\l\r} 
+ d D_\m F^{\m\l} D_\n F^\n{}_\l \biggr)\biggr] \cr
{}&{}& \sectaa \cr }
$$
Here, $a$,$b$,$c$,$d$ are constants of $O(\a)$ and $m$ is the electron mass.
The notable feature is the direct coupling of the electromagnetic field
to the curvature. This is an effective violation of the strong equivalence
principle, which states that dynamical laws should be the same in the
local inertial frames at each point in spacetime.  More precisely, this
requires the coupling of electromagnetism to gravity to be through the
connections only, independent of curvature. Eq.\sectaa ~shows that while this
principle may be consistently imposed at the classical level, it is necessarily
violated in quantum electrodynamics. 

Standard geometric optics methods applied to this effective action
results in a modified light cone \refs{\DH,\Sthree}:
\eqn\sectab{
k^2 ~-~{8\pi\over m^2}(2b+4c) T_{\m\l} k^\m k^\l ~+~{8c\over m^2} 
C_{\m\n\l\r} k^\m k^\l a^\n a^\r ~~=~~0
}
where $k^\m$ is the wave vector and $a^\m$ is the polarisation vector.
In the second term, we have replaced the Ricci tensor by the 
energy-momentum tensor using the Einstein equations. This emphasises
that this contribution is related to the presence of matter;
indeed this takes the same form as the subluminal corrections to
the speed of light in other scenarios, such as background magnetic
fields, finite temperature, etc.\refs{\LPT,\Gies}. 
The uniquely gravitational term involving the Weyl tensor
depends explicitly on $a^\m$ and gives a polarisation dependence of
the speed of light (gravitational birefringence). Moreover, 
it is readily seen that this term changes sign for the two physical,
transverse polarisations so that for vacuum (Ricci flat) spacetimes,
if one photon polarisation is subluminal, the other is necessarily
superluminal. This property, and many others, are most clearly seen
by rewriting the light cone condition \sectab ~in Newman-Penrose 
formalism \refs{\Sthree}.

The effective action \sectaa ~is, however, only the first term in a derivative
expansion, with higher order terms in $O({D\over m})$ omitted. The 
corresponding light-cone condition \sectab ~is therefore valid {\it a priori}
only in the low frequency approximation. The modified light velocity derived
from it is the phase velocity $\vp \equiv {\w\over|\ul k|}$ (where
$k_\m=(\w,\ul k)$ in a local inertial frame) at $\w \sim 0$.
In order to discuss the obvious issues concerning causality, however, the
relevant `speed of light' is not $\vp$ at $\w \sim 0$ but $\vp$ in the
high frequency limit $\w \rta \infty$. (We discuss this point carefully
in section 2.) In order to address causality, therefore, we first need to 
establish the effective light cone condition for high frequencies, which
itself involves finding the `high frequency limit' of the effective 
action \sectaa.

We have recently evaluated the effective action for QED in curved spacetime to
all orders in $O({D\over m})$, keeping terms of the form $RFF$ as in \sectaa, 
that is, the terms relevant to photon propagation to lowest order in 
$O({R\over m^2})$, i.e.~$O({\l_c^2\over L^2})$. The derivation and full details of
this result are presented in ref.\refs{\Sfive}. Here, we generalise the geometric 
optics derivation of photon propagation using this new effective action
and discuss what we can learn about the critical high frequency behaviour
of the phase velocity. The question we wish to answer is whether the
phenomenon of superluminal photon propagation is a curiosity of the 
low frequency approximation or whether it persists at high frequency, forcing
us to confront the serious implications for causality associated with
faster than light motion. In the end, our results are intriguing but not
as yet conclusive. It appears that further field-theoretic developments may
be needed to give a final resolution of the nature of dispersion in
gravitational fields.

The paper is presented as follows. 
In order to clarify exactly what we mean by the ``speed of light''
and why it is $\vp(\infty)$ which controls the causal behaviour of the
theory, we review various definitions and theorems concerning the
propagation of light in section 2. In section 3, we review the fundamentals
of geometric optics and its application to the Drummond-Hathrell action,
clarifying some subtle points arising from earlier work. Our new result
for the QED effective action in curved spacetime valid to all orders
in the derivative expansion is presented in section 4, including
some technical formulae for form factors.
The implications for photon propagation in the high frequency limit
are described in sections 5 and 6, leading to the apparent prediction
that high frequency superluminal velocities are possible in certain
spacetimes. This preliminary conclusion is challenged in section 7,
where we compare our gravitational analysis with the closely related
problem of photon propagation in a background magnetic field.
Section 8 summarises our final conclusions.  

\vskip0.7cm

\newsec{The ``Speed of Light''}

Our fundamental interest is in whether the fact that the phase velocity at
low frequencies can be superluminal (and therefore imply motion backwards
in time in a class of local inertial frames) is in contradiction with
established notions of causality. It has been discussed elsewhere that 
while in special relativity\foot{To be precise we mean here: ``in
Minkowski spacetime with no boundaries''. An analysis of causality related
to the phenomenon of superluminal propagation between Casimir plates
\refs{\Scharn,\Bart} has recently been given in ref.\refs{\LSV} 
(see also \refs{\LSVtwo,\BLV}).}
superluminal motion necessarily implies a causal paradox, in general 
relativity this is not necessarily so \refs{\DH,\Sthree}.
The key question is whether a spacetime which is {\it stably causal}
(see ref.\refs{\HE}, Proposition 6.4.9 for a precise definition)
with respect to the original metric remains stably causal with respect
to the effective metric defined by the modified light cones. 
Essentially, this means that the spacetime should still admit a foliation
into a set of hypersurfaces which are spacelike according to the 
effective light cones. This is an interesting global question, which
is beyond the scope of this paper. It seems entirely possible, however,
that the Drummond-Hathrell modifications to the light cones should
not destroy stable causality.

Before addressing such issues, however, we need to be clear what exactly we 
mean by the ``speed of light'' which determines the light cones to be used in
determining the causal structure of spacetime. In this section, we therefore
review briefly some basic definitions and results from classical optics
in order to motivate our subsequent analysis.

A particularly illuminating discussion of wave propagation in a simple
dispersive medium is given in the classic work by Brillouin \refs{\Brill}. 
This considers propagation of a sharp-fronted pulse of waves in a medium 
with a single absorption band, with refractive index $n(\w)$:
\eqn\sectba{
n^2(\w) = 1 - {a^2\over \w^2 -\w_0^2 + 2i\w\r}
}
where $a,\r$ are constants and $\w_0$ is the characteristic frequency
of the medium. Five\foot{In fact, if we take into account the distinction
discussed in section 3 between the phase velocity and the {\it ray velocity}
$v_{\rm ray}$, and include the fundamental speed of light constant $c$
from the Lorentz transformations, we arrive at 7 distinct definitions of
``speed of light''.}
distinct velocities are identified: the {\it phase
velocity} $\vp = c{\w\over{|\ul k|}} = \Re{1\over n(\w)}$, 
{\it group velocity} $v_{\rm gp} = {d\w\over d|\ul k|}$,
{\it signal velocity} $v_{\rm sig}$, {\it energy-transfer velocity} 
$v_{\rm en}$ and {\it wavefront velocity} $\vf$, with precise 
definitions related to the behaviour of contours and saddle points in the
relevant Fourier integrals in the complex $\w$-plane.
Their frequency dependence is illustrated in Fig.~1.

\vskip0.2cm
\centerline{
{\epsfxsize=8cm\epsfbox{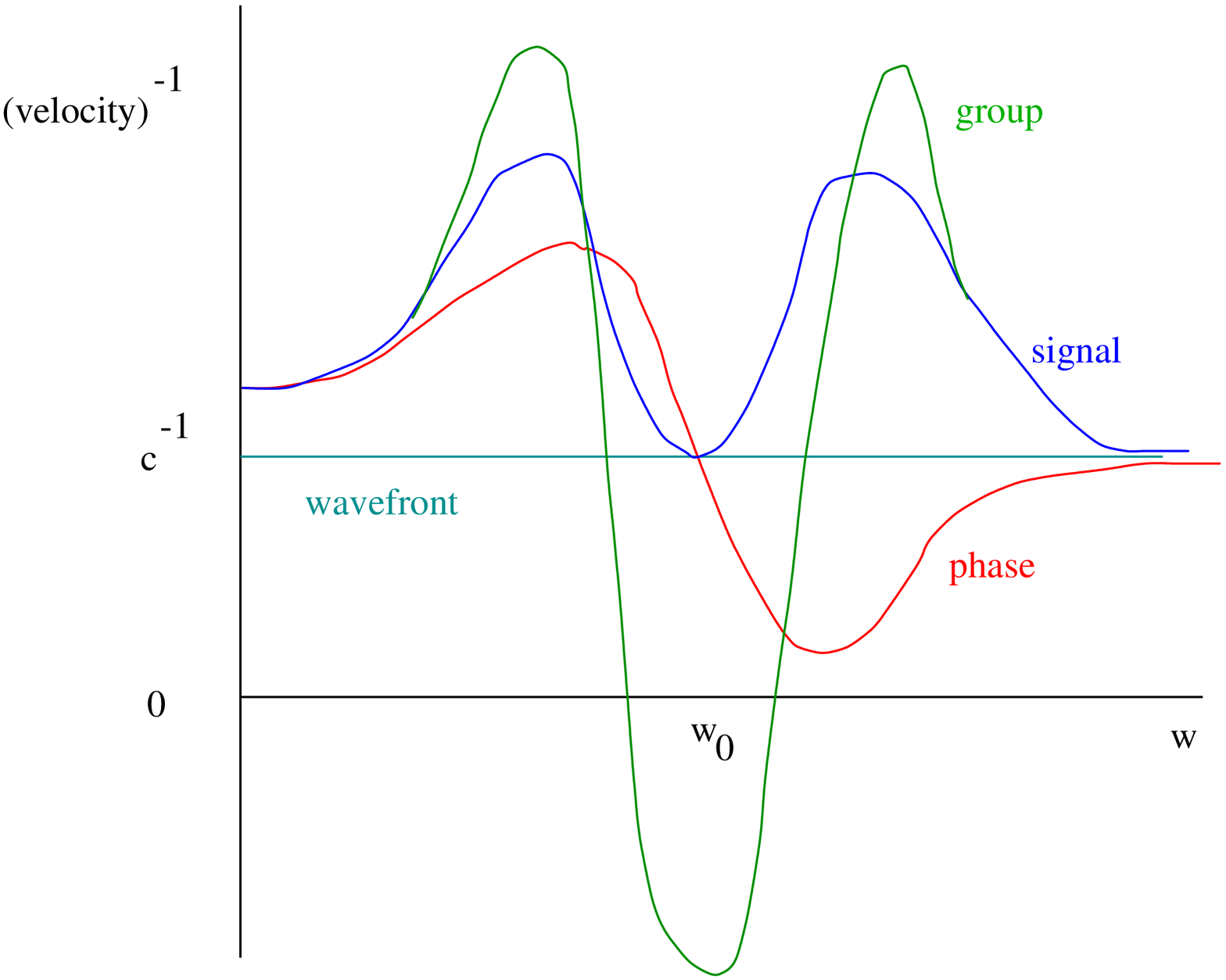}}}
\noindent{\eightpoint Fig.1~~Sketch of the behaviour of the phase, group and 
signal velocities with frequency in the model described by the refractive
index (2.1). The energy-transfer velocity (not shown) is always 
less than $c$ and becomes small near $\w_0$. The wavefront speed is 
identically equal to $c$.}
\vskip0.2cm

\noindent As the pulse propagates, the first disturbances to arrive are
very small amplitude waves, `frontrunners', which define the wavefront
velocity $\vf$. These are followed continuously by waves with amplitudes
comparable to the initial pulse; the arrival of this part of the complete
waveform is identified in \refs{\Brill} as the signal velocity $v_{\rm sig}$.
As can be seen from Fig.~1, it essentially coincides with the more familiar
group velocity for frequencies far from $\w_0$, but gives a much more
intuitively reasonable sense of the propagation of a signal than the group 
velocity, whose behaviour in the vicinity of an absorption band is
relatively eccentric.\foot{Notice that it is the group velocity which is 
measured in quantum optics experiments which find light speeds of
essentially zero \refs{\Hau} or many times $c$ \refs{\WKD}. A particularly
clear description in terms of the effective refractive index is given
in \refs{\Hau}.}
As the figure also makes clear, the phase velocity itself simply does not
represent a `speed of light' relevant for considerations of signal
propagation or causality.

The appropriate velocity to define light cones and causality is in fact
the wavefront velocity $\vf$. (Notice that in Fig.~1, $\vf$ is a constant,
equal to $c$, independent of the frequency or details of the absorption
band.) This is determined by the boundary between the regions of zero
and non-zero disturbance (more generally, a discontinuity in the first
or higher derivative of the disturbance field) as the pulse propagates.
Mathematically, this definition of wavefront is identified with the 
characteristics of the partial differential equation governing the 
wave propagation \refs{\CH}. Our problem is therefore to determine the velocity
associated with the characteristics of the wave operator derived from the
modified Maxwell equations of motion appropriate to the new effective 
action. 

A very complete and rigorous discussion of the wave equation in curved
spacetime is given in the monograph by Friedlander \refs{\Fried}, in which 
it is proved (Theorem 3.2.1) that the characteristics are simply the 
null hypersurfaces of the spacetime manifold, in other words that the 
wavefront always propagates with the fundamental speed $c$. However, this 
discussion assumes the standard form of the (gauge-fixed) Maxwell wave 
equation (cf.~ref.\refs{\Fried}, eq.(3.2.1)) and explicitly does {\it not} 
cover the modified wave equation derived from the action \sectaa, precisely 
because of the extra curvature couplings. 

Instead, the key result for our purposes, which allows a derivation
of the wavefront velocity, is derived by Leontovich \refs{\Leon}.
In this paper, an elegant proof is presented for a very general set of 
PDEs that the wavefront velocity associated with the characteristics is 
identical to the $\w\rta\infty$ limit of the phase velocity, i.e.
\eqn\sectbb{
\vf = \lim_{\w\rta \infty}{\w\over|\ul k|} = \lim_{\w\rta \infty}\vp(\w)
}
This proof appears to be of sufficient generality to apply to our
discussion of photon propagation using the modified effective action
of section 4.

The wavefront velocity in a gravitational background is therefore
not given {\it a priori} by $c$. Taking vacuum polarisation into account,
there is no simple non-dispersive medium corresponding to the 
vacuum of classical Maxwell theory in which the phase velocity 
represents a true speed of propagation; in curved spacetime QED, even
the vacuum is dispersive.
In order to discuss causality, we therefore have to extend the original
Drummond-Hathrell results for $\vp(\w \sim 0)$ to the high frequency
limit $\vp(\w\rta\infty)$, as already emphasised in ref.\refs{\DH}. 
This is why the effective action \sectaa ~to lowest order in the derivative 
expansion is not sufficient and we require the all-orders effective
action of section 4. 

A final twist emerges if we write the standard dispersion relation
for the refractive index $n(\w)$ in the limit $\w\rta\infty$:
\eqn\sectbc{
n(\infty) = n(0) - {2\over\pi} \int_0^\infty {d\w\over\w} \Im n(\w)
}
For a standard dispersive medium, $\Im n(\w) > 0$, which implies that
$n(\infty) < n(0)$, or equivalently $\vp(\infty) > \vp(0)$.
Evidently this is satisfied by Fig.~1. The key question though is 
whether the usual assumption of positivity of $\Im n(\w)$ holds 
in the present situation of the QED vacuum in a gravitational field.
If so, then (as already noted in ref.\refs{\DH}) the superluminal
Drummond-Hathrell results for $\vp(0)$ would actually be {\it lower
bounds} on the all-important wavefront velocity $\vp(\infty)$.
However, it is not so clear that positivity of $\Im n(\w)$ is reliable
in the gravitational context. Indeed it has been explicitly
criticised by Dolgov and Khriplovich in refs.\refs{\DK,\Khrip},
who point out that since gravity is an inhomogeneous medium in which beam
focusing as well as diverging can happen (see next section), a growth 
in amplitude corresponding to $\Im n(\w) < 0$  is possible. 
It therefore seems best to set the dispersion relation \sectbc ~aside 
for the moment and concentrate instead on
a direct attempt to determine $\vp(\infty)$. This is the goal of this 
paper.

\vskip0.7cm

\newsec{Low Frequency Photon Propagation}

The simplest way to deduce the light-cone condition implied by the
effective action is to use geometric optics. In this section, we review
the approach introduced in ref.\refs{\DH} (see also refs.{\Sone,\Sthree})
emphasising some points which will be important later.

In geometric optics (see \refs{\SEF} for a thorough discussion) the 
electromagnetic field is written as the product
of a slowly-varying amplitude and a rapidly-varying phase, i.e.
\eqn\sectca{
F_{\m\n} = \Re~(f_{\m\n} + i \e h_{\m\n} + \ldots)e^{{i\over\e}\vartheta}
}
Here, $\e$ is a parameter introduced purely as a book-keeping device
to keep track of the relative order of magnitude of terms. The field 
equations and Bianchi identities are solved order by order in $\e$.

The wave vector is defined as the derivative of the phase, i.e.
$k_\m = \pl_\m \vartheta$.
The leading order term in the Bianchi identity $D_{[\l} F_{\m\n]}=0$
is of $O({1\over\e})$ and constrains $f_{\m\n}$ to have the form
\eqn\sectcb{
f_{\m\n} = k_\m {\cal A}_\n - k_\n {\cal A}_\m
}
where ${\cal A}_\m = {\cal A} a_\m$. ${\cal A}$ represents the
amplitude while $a^\m$ (normalised so that $a^\m a_\m = -1$) specifies 
the polarisation. For physical polarisations, $k_\m a^\m = 0$.

Conventional curved spacetime QED is based on the usual Maxwell action,
so the equation of motion is simply 
\eqn\sectcc{
D_\m F^{\m\n} = 0
}
At leading order, $O({1\over\e})$, this becomes
\eqn\sectccc{
k_\m f^{\m\n} = 0
}
and since this implies
\eqn\sectcd{
k^2 a^\n = 0
}
we immediately deduce that $k^2=0$, i.e.~$k^\m$ is a null vector.
Then, from its definition as a gradient, we see 
\eqn\sectce{
k^\m D_\m k^\n ~=~ k^\m D^\n k_\m ~=~ {1\over2} D^\n k^2 ~=~ 0
}
Light rays (photon trajectories) are defined as the integral curves 
of $k^\m$, i.e.~the curves $x^\m(s)$ where ${dx^\m \over ds} = k^\m$.
These curves therefore satisfy
\eqn\sectcf{
0 ~=~ k^\m D_\m k^\n 
~=~ {d^2 x^\n \over ds^2} + \C^\n_{\m\l} {dx^\m \over ds}{dx^\l \over ds}
}
which is the geodesic equation. So in the conventional theory, light rays
are null geodesics.

The subleading, $O(1)$, term in the equation of motion gives
\eqn\sectcfa{
k^\m D_\m {\cal A}^\n = -{1\over2}(D_\m k^\m){\cal A}^\n
}
which decomposes into
\eqn\sectfcb{
k^\m D_\m a^\n = 0
}
and
\eqn\sectfcc{
k^\m D_\m (\ln {\cal A}) = - {1\over2}D_\m k^\m 
}
The first shows that the polarisation vector is parallel transported
along the null geodesic rays while the second, whose r.h.s.~is simply minus
the optical scalar $\theta$, shows how the amplitude varies as the beam of 
rays focuses or diverges.

We now apply the same methods to the modified effective action \sectaa.
This gives rise to a new equation of motion 
which, under the approximations listed below, simplifies to:
\eqn\sectcga{
D_\m F^{\m\n} ~-~ {1\over m^2} \biggl[2b R_{\m\l}D^\m F^{\l\n}
+ 4c R_\m{}^\n{}_{\l\r}D^\m F^{\l\r}\biggr]~~=~~0
}
Here, we have neglected derivatives of the curvature tensor, which would  
be suppressed by powers of $O(\l/L)$, where $\l$ is the photon wavelength and 
$L$ is a typical curvature scale, and we have omitted the new contributions 
involving $D_\m F^{\m\n}$~: since this term is already $O(\a)$ using the
equations of motion, these contributions only affect the light cone
condition at $O(\a^2)$ and must be dropped for consistency.
Making the standard geometric optics assumptions described above,
we then find the new light cone condition:
\eqn\sectcg{
k_\m f^{\m\n} ~-~ {1\over m^2} \biggl[2b R^\m{}_\l k_\m f^{\l\n}
+ 4c R^{\m\n}{}_{\l\r} k_\m f^{\l\r} \biggr] ~~=~~0
}
Eq.\sectcg ~can now be rewritten as an equation for the polarisation vector
$a^\m$, and re-expressing in terms of the Weyl tensor we find
\eqn\sectch{
k^2 a^\n ~-~ {(2b+4c)\over m^2}~ R_{\m\l} \bigl(k^\m k^\l a^\n
- k^\m k^\n a^\l \bigr) ~-~ {8c\over m^2} ~ 
C_\m{}^\n{}_{\l\r} k^\m k^\l a^\r ~~=~~0
}
The solutions of this equation describe the propagation
for a photon of wave vector $k_\m$ and polarisation $a^\m$. Contracting
with $a_\n$, we find the effective light cone
\eqn\sectci{
k^2 ~-~{(2b+4c)\over m^2} R_{\m\l} k^\m k^\l ~+~{8c\over m^2} 
C_{\m\n\l\r} k^\m k^\l a^\n a^\r ~~=~~0
}
from which \sectab ~follows immediately.

Notice that in the discussion of the free Maxwell theory, we did not need to
distinguish between the photon momentum  $p^\m$, i.e.~the tangent vector 
to the light rays, and the wave vector $k_\m$ since they were simply related
by raising the index using the spacetime metric, $p^\m = g^{\m\n}k_\n$. 
In the modified theory, 
there is an important distinction. The wave vector, defined as the derivative
of the phase, is a covariant vector or 1-form, whereas the photon momentum/
tangent vector to the rays is a true contravariant vector. The relation
is non-trivial. In fact, when as in \sectci ~we can write the light cone
condition for the wave vector as the homogeneous form
\eqn\sectcj{
{\cal G}^{\m\n}k_\m k_\n = 0
}
we should define the corresponding
`momentum' as 
\eqn\sectck{
p^\m = {\cal G}^{\m\n}k_\n
}
and the light rays as curves $x^\m(s)$ where ${dx^\m\over ds} = p^\m$.
This definition of momentum satisfies 
\eqn\sectcl{
G_{\m\n} p^\m p^\n = {\cal G}^{\m\n}k_\m k_\n = 0
}
where $G \equiv {\cal G}^{-1}$ therefore defines a new effective metric
which determines light cones mapped out by the geometric optics light rays.
(Indices are always raised or lowered using the true metric $g_{\m\n}$.)
The {\it ray velocity} $v_{\rm ray}$ corresponding to the momentum $p^\m$, 
which is the velocity with which the equal-phase surfaces advance, is given by
\eqn\sectcm{
v_{\rm ray} = {|\ul p|\over p^0} = {d |\ul x|\over dt}
}
along the ray, and is in general different from the phase velocity
\eqn\sectcn{
\vp = {k^0\over|\ul k|}
}

A nice example of this is given in ref.\refs{\LSV}, which analyses certain
aspects of superluminal propagation in Minkowski spacetime with Casimir plates.
The discrepancy between $\vp$ and $v_{\rm ray}$ (called $c_{\rm light}$ in
\refs{\LSV}) in that example is due to a difference between the direction of
propagation along the rays and the wave 3-vector. Otherwise, it follows 
directly from \sectcl ~that $v_{\rm ray}$ and $\vp$ are identical.

Recognising the distinction between $p$ and $k$ also clarifies a potentially
confusing point in the important example of propagation in a FRW 
spacetime \refs{\DH}.
Since the FRW metric is Weyl flat, the modified light cone condition \sectab
~reads simply
\eqn\sectco{
k^2 = \zeta~ T_{\m\n} k^\m k^\n 
}
where $\zeta = {8\pi\over m^2}(2b+4c)$ and the energy-momentum tensor is
\eqn\sectcp{
T_{\m\n} = (\r + P)n_\m n_\n - P g_{\m\n}
}
with $n^\m \equiv (e_t)^\m$ specifying the time direction in a comoving
orthonormal frame. $\r$ is the energy density and $P$ is the
pressure, which in a radiation-dominated era are related by $\r - 3P = 0$.
The phase velocity is independent of polarisation and is found to be
superluminal:
\eqn\sectcq{
\vp = {k^0\over |\ul k|} = 1 + {1\over2}\zeta (\r + P)
}
At first sight, this looks surprising given that $k^2 > 0$. 
However, if instead we consider the momentum along the rays,
we find
\eqn\sectcr{
p^2 = g_{\m\n}p^\m p^\n = -\zeta (\r + P) (k^0)^2
}
and 
\eqn\sectcs{
v_{\rm ray} = {|\ul p|\over p^0} = 1 + {1\over2}\zeta(\r + P)
}
The effective metric $G = {\cal G}^{-1}$ is
\eqn\sectcs{
G_{\m\n} = \left(\matrix{1 + \zeta(\r+P) &0&0&0\cr
0&-1&0&0\cr 0&0&-1&0\cr 0&0&0&-1\cr}\right)
}
In this case, therefore, we find equal, superluminal
velocities $\vp = v_{\rm ray}$ and $p^2 < 0$ is manifestly
spacelike as required. 

In the radiation dominated era, 
where $\r(t) = {3\over 32\pi} t^{-2}$, we have
\eqn\sectct{
\vp = 1 + {1\over 16\pi}\zeta~ t^{-2}
}
which, as already observed in \refs{\DH}, increases towards the early
universe. Although this expression is only reliable in the perturbative
regime where the correction term is small, it is intriguing that QED
predicts a rise in the speed of light in the early universe. It is a matter
of speculation whether this superluminal effect persists for high
curvatures ($L\sim\l_c$) and whether it could be important in the context
of the horizon problem.

\vskip0.7cm

\newsec{The Effective Action for QED in a Gravitational Field}

The effective action presented in this section has recently been derived 
\refs{\Sfive} by adapting the more general background field calculations
of \refs{\BGZV}, this latter paper being the culmination of the theoretical
development described in \refs{\BV}. The result is
an effective action which incorporates the one-loop vacuum polarisation
contributions to the photon propagator in an arbitrary, weak gravitational
field. We therefore keep terms of type $RFF$, i.e.~quadratic in the 
electromagnetic field but only of first order in the curvature. This   
neglects terms of higher orders in $O({\l_c^2\over L^2})$. However, the new 
feature compared to the Drummond-Hathrell action \sectaa ~is that terms
to all orders in derivatives are kept. This allows a discussion of the 
frequency dependence of the modifications to photon propagation.

The effective action is given by
\eqn\sectda{
\C = \C_{(0)} + \ln {\rm det} S(x,x')
}
where $\C_{(0)}$ is the free Maxwell action and $S(x,x')$ is the Green
function of the Dirac operator in the background gravitational field,
i.e.
\eqn\sectdb{
\bigl(i\Dslash  - m\bigr) S(x,x') = {i\over\sqrt{ -g}} \d(x,x')
}

In fact it is more convenient to introduce the scalar Green function
$G(x,x')$ defined by
\eqn\sectdc{
S(x,x') = \bigl(i\Dslash  + m\bigr) G(x,x')
}
so that 
\eqn\sectdd{
\Bigl( D^2 + ie\s^{\m\n} F_{\m\n} - {1\over 4} R + m^2\Bigr) G(x,x')~~=~~
-{i\over\sqrt{ -g}} \d(x,x')
}
Then we evaluate $\C$ from the heat kernel, or proper time, representation
\eqn\sectde{
\C ~~=~~ \C_{(0)} ~-~{1\over2}\int_0^\infty {ds\over s} ~e^{-i m^2 s}~ 
{\rm Tr} {\cal G}(x,x';s)
}
where
\eqn\sectdf{
{\cal D}{\cal G}(x,x';s) = i {\pl\over\pl s}{\cal G}(x,x';s)
}
with ${\cal G}(x,x';0) = G(x,x')$. Here, ${\cal D}$ is the differential 
operator in eq.\sectdd ~at $m=0$.

The details of the derivation of the effective action are given in 
ref.\refs{\Sfive} and here we simply quote the result: 
\eqnn\sectdg
$$\eqalignno{
\C  = \int dx \sqrt{-g} \biggl[-{1\over4}F_{\m\n}F^{\m\n}~
&+~{1\over m^2}\Bigl(D_\m F^{\m\l}~ \or{G_0}~ D_\n F^\n{}_{\l} \cr
&~~~~~~~+~\or{G_1}~ R F_{\m\n} F^{\m\n}~ 
+~\or{G_2}~ R_{\m\n} F^{\m\l}F^\n{}_{\l}~
+~\or{G_3}~ R_{\m\n\l\r}F^{\m\n}F^{\l\r} \Bigr)\cr
&+~{1\over m^4}\Bigl(\or{G_4}~ R D_\m F^{\m\l} D_\n F^\n{}_{\l} \cr
&~~~~~~~+~\or{G_5}~ R_{\m\n} D_\l F^{\l\m}D_\r F^{\r\n}~
+~\or{G_6}~ R_{\m\n} D^\m F^{\l\r}D^\n F_{\l\r} \cr
&~~~~~~~+~\or{G_7}~ R_{\m\n} D^\m D^\n F^{\l\r} F_{\l\r}~
+~\or{G_8}~ R_{\m\n} D^\m D^\l F_{\l\r} F^{\r\n} \cr
&~~~~~~~+~\or{G_9}~ R_{\m\n\l\r} D_\s F^{\s\r}D^\l F^{\m\n}~\Bigr)~~ 
\biggr] \cr 
{}&{}& \sectdg \cr }
$$
In this formula, the $\or{G_n}$ ($n\ge 1$) are form factor functions of 
three operators:
\eqn\sectdh{
\or{G_n} \equiv G_n\Bigl({D_{(1)}^2\over m^2}, {D_{(2)}^2\over m^2}, 
{D_{(3)}^2\over m^2}\Bigr)
}
where the first entry ($D_{(1)}^2$) acts on the first following term
(the curvature), etc. $\or{G_0}$ is similarly defined as a single variable 
function. The ${G_n}$ are themselves expressed as proper time integrals:
\eqn\sectdi{
G_n(x_1,x_2,x_3) ~~=~~-{1\over2}{\a\over\pi}~\int_0^\infty{ds\over s}
e^{-is} (is)^p ~g_n(-isx_1, -isx_2, -isx_3)
}
where $p=1$ for $n=0,\ldots 3$ and $p=2$ for $n=4,\ldots 9$, and we have
rescaled $s$ by a factor of $m^2$ so as to be a dimensionless variable. 

A crucial feature of this form of the effective action is that it is
{\it local}, in the sense that the form factors $\or{G_n}$ have an
expansion in positive powers of the $D_{(i)}^2$. This depends on making
the choice of basis operators above, in contrast to the original
form quoted in ref.\refs{\BGZV}. 

The values of $G_n(0,0,0)$ for $n=0,\ldots 3$ reproduce the Drummond-Hathrell
results. For these, we have $G_n(0,0,0) = -{1\over2}{\a\over\pi}~g_n(0,0,0)$, 
where:
\eqnn\sectdj
$$\eqalignno{
a &= -{1\over2}{\a\over\pi} ~g_1(0,0,0) ~=~ 
-{1\over 144}{\a\over\pi}~~~~~~~~~~~~~~~~~~~~
b = -{1\over2}{\a\over\pi} ~g_2(0,0,0) ~=~ {13\over 360}{\a\over\pi}\cr
c &= -{1\over2}{\a\over\pi} ~g_3(0,0,0) ~=~ 
-{1\over 360}{\a\over\pi}~~~~~~~~~~~~~~~~~~~~
d = -{1\over2}{\a\over\pi} ~g_0(0) ~=~ -{1\over 30}{\a\over\pi}\cr
{}&{}& \sectdj \cr }
$$

Explicit analytic forms for all the form factors $\or{G_n}$ are known
and are given in detail in ref.\refs{\Sfive}. Here, we simply quote
the expressions for the form factors relevant
for Ricci flat spaces, $g_3(x_1,x_2,x_3)$ and $g_9(x_1,x_2,x_3)$.
Moreover, since terms involving derivatives acting on the curvature
are identified as higher order in ${\l_c^2\over L^2}$ in the present context, 
we restrict to the special case $x_1=0$.
Then, from ref.\refs{\Sfive}, we find:
\eqnn\sectdk
$$\eqalignno{
g_3(0,x_2,x_3) ~~=~~&-~F(0,x_2,x_3) {1\over\D}\Bigl[4+\bigl(1+{1\over4}
(x_2+x_3)\bigr)(x_2+x_3)\Bigr] \cr
&+~f(x_2){1\over\D}\Bigl[{3x_2-x_3\over2x_2} + {1\over4}(x_2-x_3)\Bigl(1-
{2(x_2+x_3)^2\over\D}\Bigr) \Bigr] \cr
&+~f(x_3){1\over\D}\Bigl[{3x_3-x_2\over2x_3} + {1\over4}(x_3-x_2)\Bigl(1-
{2(x_2+x_3)^2\over\D}\Bigr) \Bigr] \cr
&+~{1\over\D}\Bigl[-1 + {1\over2}\Bigl({x_3\over x_2} + {x_2\over x_3} + 
x_2 + x_3 \Bigr)\Bigr] ~
+~{1\over12}\Bigl({1\over x_2} + {1\over x_3}\Bigr) ~+~ {1\over4} \Bigl(
{1\over x_2^2} + {1\over x_3^2}\Bigr) \cr
{}&{}& \sectdk \cr }
$$
and 
\eqnn\sectdl
$$\eqalignno{
g_9(0,x_2,x_3) ~~=~~&-~F(0,x_2,x_3) {1\over\D}\bigl[4 + x_2 + x_3\bigr] \cr
&+~f(x_2)\Bigl[{2\over \D^2}(x_2^2 - x_3^2) - {1\over 2x_2^2} - 
{2\over x_2^3}\Bigr] \cr
&+~f(x_3) \Bigl[-{2\over\D^2}(x_2^2 - x_3^2)\Bigr]~
+~f'(x_2)\Bigl({1\over x_2^2} + {1\over 2x_2}\Bigr) ~-{2\over\D} + 
{1\over 3x_2^2} + {2\over x_2^3} \cr
{}&{}& \sectdl \cr }
$$
where $\D = (x_2-x_3)^2$. It can be checked that all the inverse powers
of $x_2$ and $x_3$ cancel leaving a finite $x_2=x_3=0$ limit, as required
for the form factors to be local.
Here,
\eqn\sectdm{
f(x) ~~=~~ \int_0^1 d\a~e^{-\a(1-\a)x}
}
and
\eqn\sectdn{
F(x_1,x_2,x_3) ~~=~~\int_{\a\ge0} d^3\a~\d(1-\a_1-\a_2-\a_3)~
e^{-\a_1 \a_2 x_3 - \a_2 \a_3 x_1 - \a_3 \a_1 x_2}
}
It will also be useful to note the simpler expression
\eqn\sectdo{ 
F(0,x_2,x_3) ~~=~~ {1\over x_2-x_3} \int_0^1 d\a~{1\over \a}~
\Bigl[ e^{-\a(1-\a)x_3} - e^{-\a(1-\a)x_2} \Bigr]
}

The behaviour of these functions is illustrated in the following plots
of $g_3(0,x_2,x_3)$ and $g_9(0,x_2,x_3)$. Notice that along the diagonals
$x_2=x_3$, both functions tend asymptotically to zero. However, if one
argument is set to zero, then the functions may tend to a finite limit.
The values at the origin are $g_3(0,0,0) = {1\over180}$ and
$g_9(0,0,0) = {5\over504}$.

\vskip0.6cm
{\epsfxsize=6cm\epsfbox{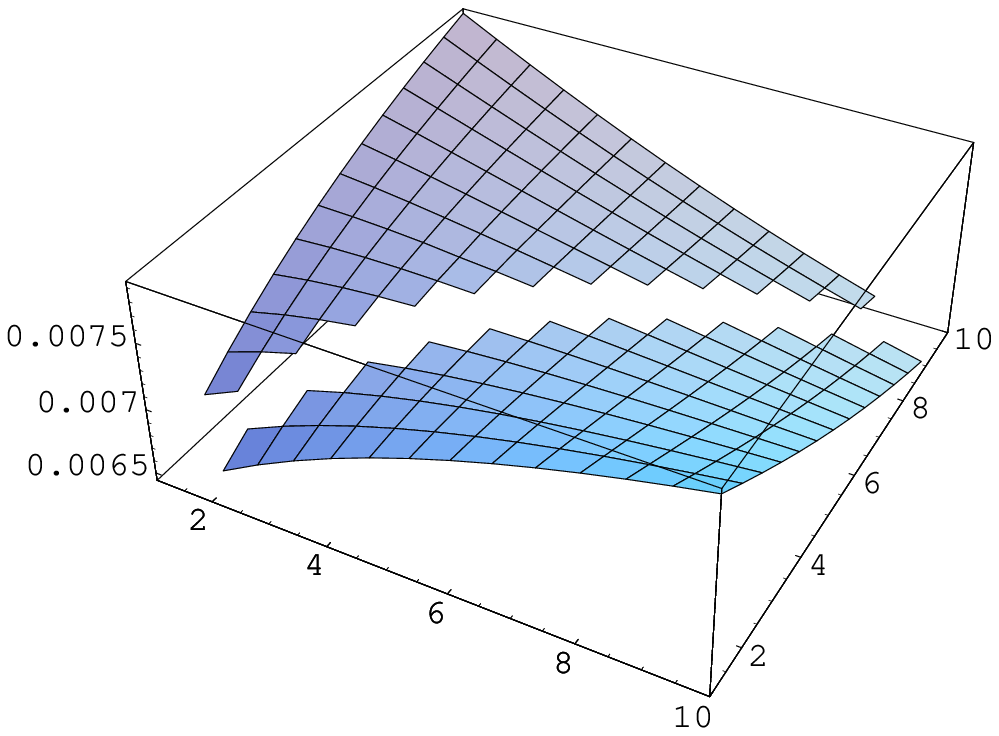}}\hskip1cm
{\epsfxsize=6cm\epsfbox{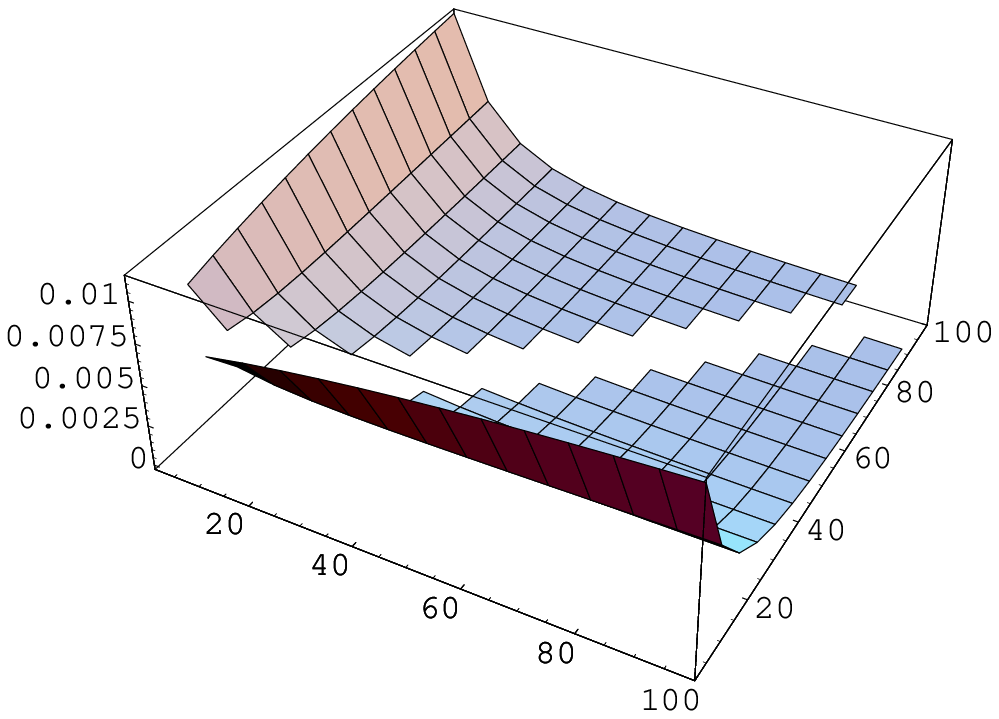}}

\noindent{\eightpoint Fig.2~~3D plots of $g_3(0,x_2,x_3)$ over different ranges.}

\vskip0.6cm
{\epsfxsize=6cm\epsfbox{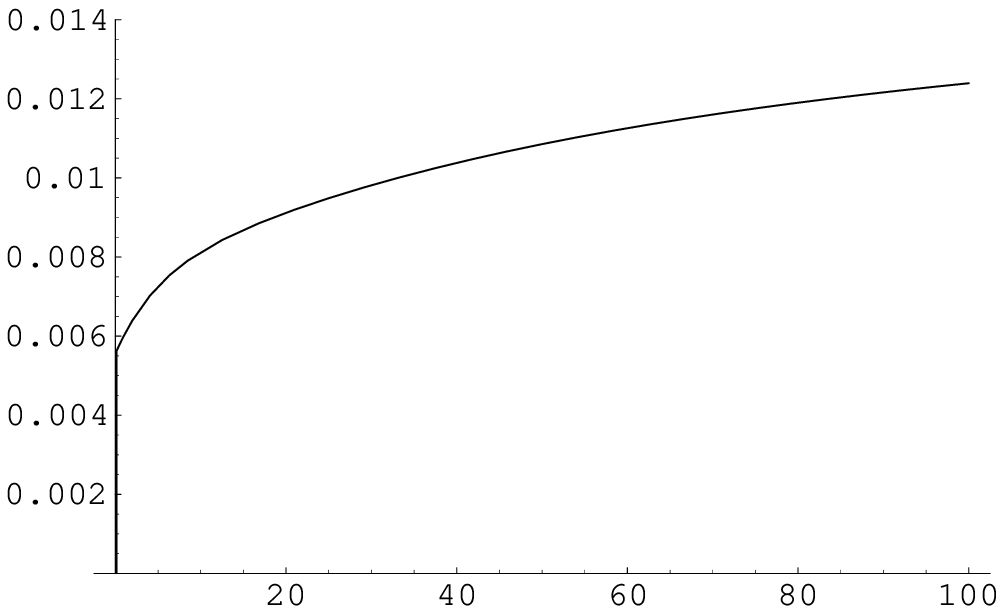}}\hskip1cm
{\epsfxsize=6cm\epsfbox{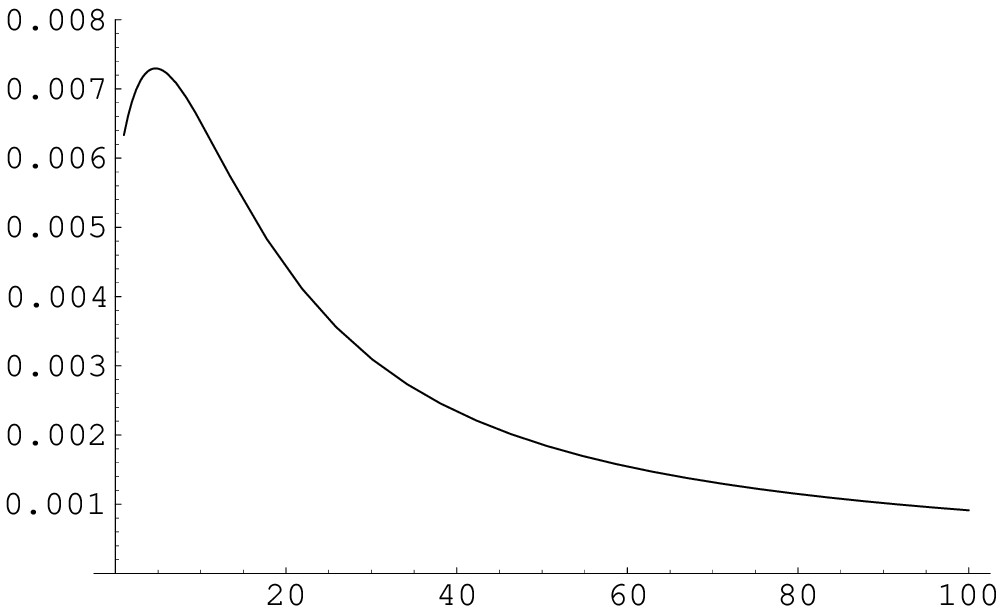}}

\noindent{\eightpoint Fig.3~~Graphs of $g_3(0,x_2,x_3)$ along the $x_2$ or $x_3$ 
axes (left) and the diagonal $x_2=x_3$ (right).}
\vskip0.3cm

\vfill\eject

\vskip0.6cm
{\epsfxsize=6cm\epsfbox{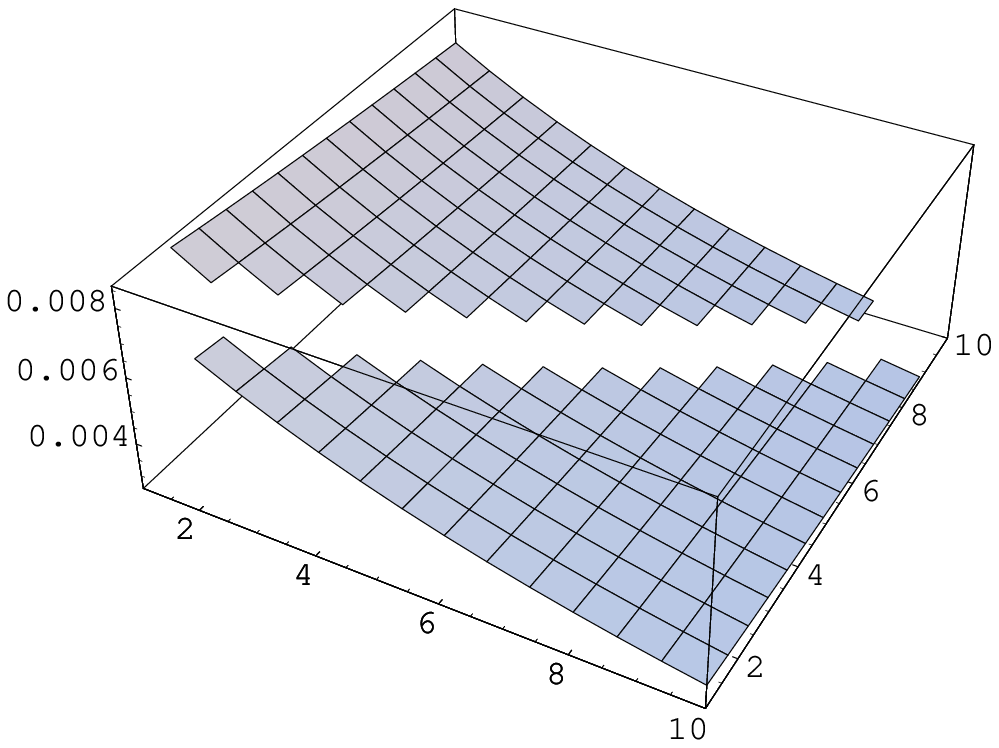}}\hskip1cm
{\epsfxsize=6cm\epsfbox{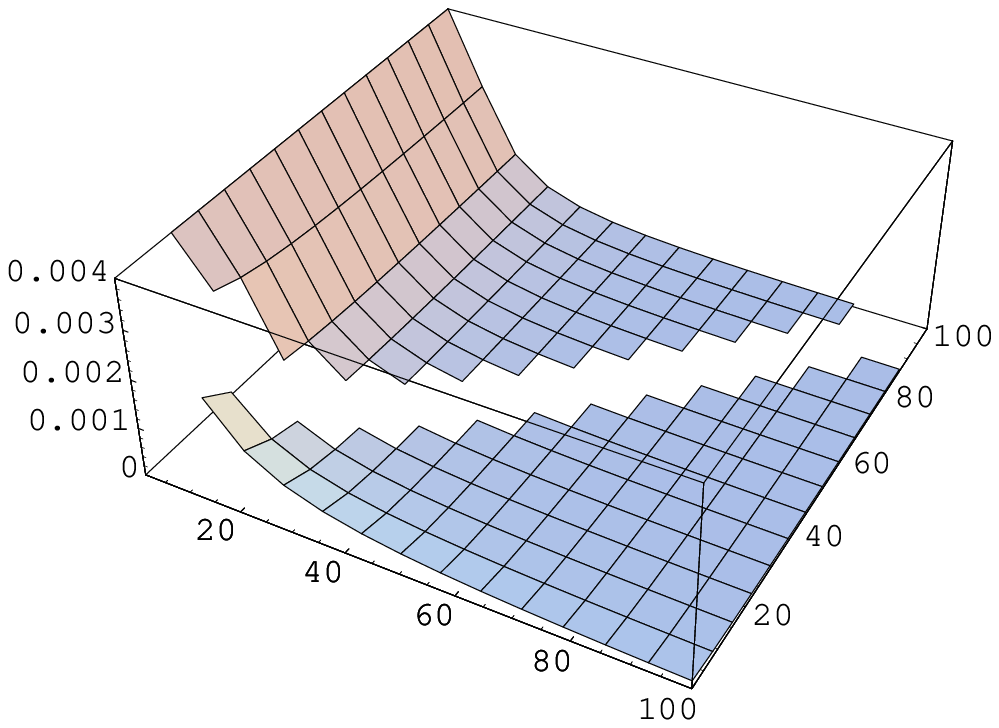}}

\noindent{\eightpoint Fig.4~~3D plots of $g_9(0,x_2,x_3)$ over different ranges.}
\vskip0.3cm

\vskip0.6cm
{\epsfxsize=6cm\epsfbox{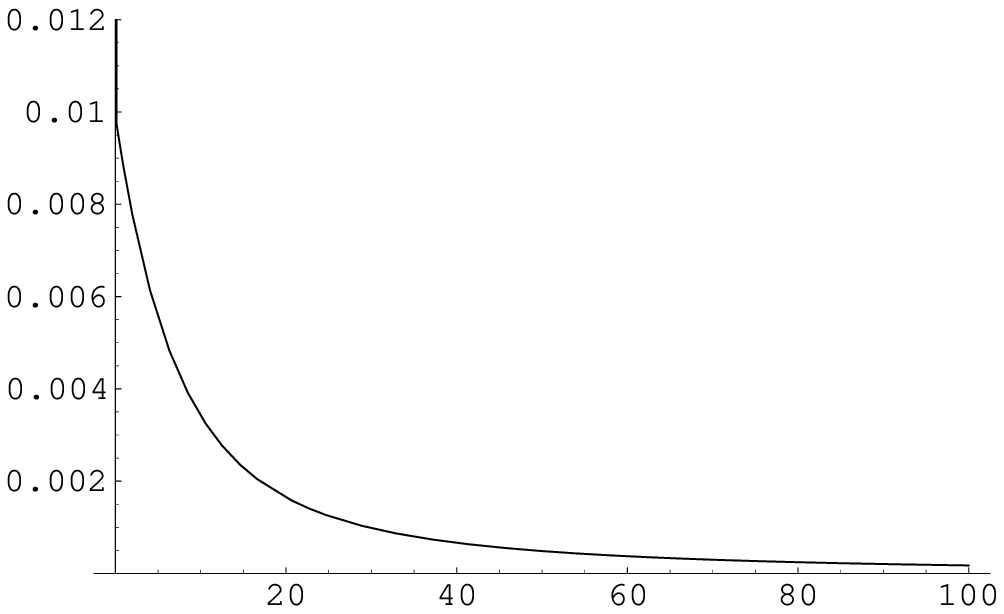}}\hskip1cm
{\epsfxsize=6cm\epsfbox{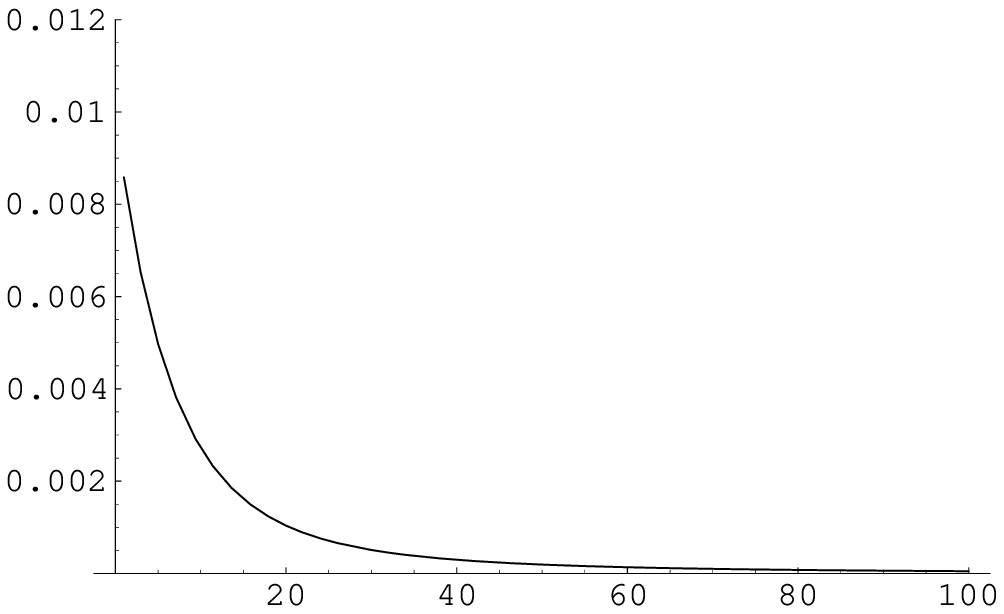}}
\vskip0.6cm
{\epsfxsize=6cm\epsfbox{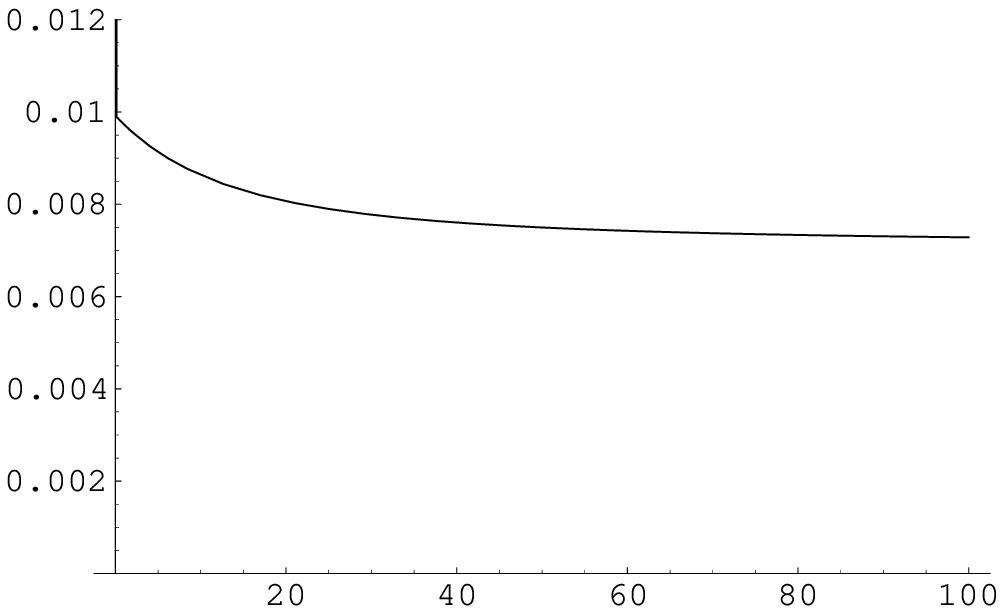}}

\noindent{\eightpoint Fig.5~~Graphs of $g_9(0,x_2,x_3)$ along the $x_2$ 
axis (top left), the diagonal $x_2=x_3$ (top right) and the $x_3$ axis (lower).}
\vskip0.3cm

\vfill\eject

\newsec{Geometric Optics with the New Effective Action}

In this section, we apply the geometric optics methods introduced
in section 3 to the effective action \sectdg. In order to simplify
the following discussion, we restrict here to the special case
of Ricci flat spacetimes. This will be sufficient to extract
the most important information, viz.~the influence of the purely
gravitational Weyl tensor contributions to high frequency photon
propagation.

It is easiest to present our results by building up from simplified
cases\foot{Some early related analysis can be found in ref.\refs{\Myers}}.
So we consider first the modifications to the 
equation of motion ${\delta\C\over\d A^\n}=0$ from the $\or{G}_3$ and 
$\or{G}_9$ terms neglecting their non-trivial derivative dependence.
We then find:
\eqn\sectea{
D_\m F^{\m\n} + {1\over m^2}\biggl[-4 G_3(0,0,0)~ 
R_\m{}^\n{}_{\l\r}D^\m F^{\l\r} ~
+~ G_9(0,0,0)~ D^2 \bigl(R_\m{}^\n{}_{\l\r}D^\m F^{\l\r}\bigr) 
\biggr] ~=~0
}
Here, we have discarded all the $O(\a)$ terms involving 
$D_\m F^{\m\n}$ for the reason explained in section 3, together with some 
terms involving derivatives of the curvature which are always suppressed. 
Notice that we are free to interchange covariant derivatives at will,
since a commutator produces a further power of curvature and therefore
a further $O({\l_c^2\over L^2})$ suppression. We have also used
the Bianchi identities for the curvature and omitted terms involving the
Ricci tensor. The important observation, however, is that
the only effect on the equation of motion of the derivatives in the 
structure of the $G_9$ term in the effective action is to produce an extra
$D^2$ in the form factor. 

We now come to the implementation of the geometric optics approximation
taking into account the non-trivial $D^2$ dependence in the form factors.
Again, we illustrate this with a simple example. Consider a term
\eqn\secteb{
{1\over m^4} \int dx \sqrt{-g}~R_{\m\n\l\r}F^{\m\n} D^2 F^{\l\r}
}
in the effective action. This gives the following contribution to the 
equation of motion for $D_\m F^{\m\n}$:
\eqn\sectec{
{1\over m^4} \biggl[
-2 \bigl(D^2 R_\m{}^\n{}_{\l\r}\bigr) D^\m F^{\l\r}
-4 \bigl(D_\s R_\m{}^\n{}_{\l\r}\bigr) D^\s D^\m F^{\l\r} 
-4 R_\m{}^\n{}_{\l\r} D^2 D^\m F^{\l\r} \biggr]
}
Compared with the basic $G_3$ term above, the relative orders of these
terms are as follows. The first is suppressed by $O({\l_c^2\over L^2})$,
since it involves extra derivatives of the curvature.
The second involves an extra derivative of the curvature tensor but a
compensating extra power of $k$ from the associated derivative acting on
the field strength, so overall this term is of relative order
$O({\l_c^2\over \l L})$. In the low frequency region $\l > \l_c$ considered
in \refs{\DH} this would be neglected, but here we are interested in
extending the results to $\l < \l_c$. 
At first glance, the third term appears to be dominant because of the 
extra two powers of $k$ coming from $D^2$ acting on $F^{\l\r}$.
However, this appears contracted as $k^2$ which is not large but 
rather zero at leading order, and $O(\a)$ in the full theory. So in fact
this term must be discarded for consistency with the perturbative expansion
in $\a$.

Returning to the second term, after substituting the geometric optics
ansatz for $F^{\l\r}$, we find that we recover the same structural form
for the light cone condition but with a factor now involving
${1\over m^4}k\cdot D R_{\m\n\l\r}$.
It is clear that after we generalise from \secteb ~to the complete form
factors, these terms will sum up and we will find the light cone modified by
functions of $k\cdot D$ acting on the curvature.

First, consider the order of magnitude of these corrections, 
$O({\l_c^2\over \l L})$. This is interesting because it is closely related 
to the condition for direct observability of the Drummond-Hathrell effect.
Consider differently polarised light propagating with a velocity
difference of $O(\a{\l_c^2\over L^2})$, as predicted by eq.\sectab,
over a time $L$ characteristic of the spacetime curvature. This produces the
biggest length difference between the rays which can be realised in 
the spacetime. To be observable, this should (as a rough order of magnitude)
be greater than the wavelength $\l$. We therefore arrive at the 
following criterion for direct observability:
\eqn\sected{
\a {\l_c^2 \over L^2} L ~>~ \l
}
that is
\eqn\sectee{
\a {\l_c^2 \over \l L} ~>~ 1
}

We see that to access the frequency range where superluminal effects
could in principle be observable, we need to satisfy the criterion
${\l_c^2 \over \l L} \gg 1$. But since, as we have just shown, this is
the parameter governing the corrections to the light cone from the
new form factor terms in the effective action, we find that  
these terms cannot be neglected as perturbatively small
but must instead be summed to all orders.
 
Implementing this strategy, it can now be shown that the final
light cone condition following from the equations of motion derived
from \sectdg ~is:
\eqn\sectef{
k^2 ~+~ {2\over m^2}~
\biggl[4~ G_3\Bigl(0, {2ik\cdot D\over m^2}, 0\Bigr)~-~
{2ik\cdot D} ~G_9\Bigl(0, {2ik\cdot D\over m^2}, 0\Bigr) \biggr]~
R_{\m\n\l\r} k^\m k^\l a^\n a^\r ~=~0
}
In the derivation, we have used the symmetry of $G_3(x_1,x_2,x_3)$ under
$x_2 \leftrightarrow x_3$ (not present in $G_9$ of course), and omitted the
$O({\l_c^2\over L^2})$ suppressed terms from the operator $D_{(1)}^2$
acting on $R_{\m\n\l\r}$.

Eq.\sectef ~is therefore the generalisation of the light cone
\sectab, in the Ricci flat case, where we extend the effective action to
include the non-trivial form factor operators $\or{G}_n$. In the light cone 
condition, these form factors reduce to single-variable functions of the 
operator $k \cdot D$ acting on the curvature tensor $R_{\m\n\l\r}$. Although
the effective parameter ${1\over m^2}k\cdot D $ is $O({\l_c^2\over\l L})$ 
and therefore not small in the region of interest, knowledge of the analytic 
expressions for the form factors should now enable these corrections to be 
exactly summed. We claim that this is the correct treatment of the equation
of motion with extra derivatives in the perturbative corrections
in a self-consistent geometric optics expansion.

In the next section, we study the condition \sectef ~in more detail
and discuss what we can learn from it about the high frequency
behaviour of the phase velocity $\vp$. Before that, we make a few comments
on the relation of our result to the analysis of Khriplovich \refs{\Khrip}.
In a very interesting contribution to the debate on dispersion and the 
Drummond-Hathrell mechanism, Khriplovich considered the general structure of 
the three-particle vertex for the scattering of an on-shell photon by a graviton.
The 3 possible Lorentz structures are shown to be equivalent to the 3 terms
in the Drummond-Hathrell action, acted on by form factors which are functions
of the invariant momenta for the photon and graviton legs. The essential 
observation of Khriplovich is that for on-shell (i.e. $k^2=0$) photons,
the form factors reduce to constants, given by the coefficients $a,b,c$ of
section 3. The conclusion is then that there is {\it no} dispersion: the
light cone shift at $\w=0$ is unchanged for all frequencies.

Clearly, this differs from our conclusions here. The reason is not entirely
clear. However, the argument in \refs{\Khrip} for the triviality of the 
form factors in the vertex looks remarkably similar to the discussion below
\sectec. In our case, we also had form factors which were functions of
$D^2$ acting on the electromagnetic field, and applying geometric optics
these at first sight would appear to give contributions only of $O(k^2)$
which would vanish. However, the more complete analysis showed that they
nevertheless can give rise to contributions involving $k\cdot D~R$ which
do not vanish. The most likely explanation for the discrepancy between our
results and ref.\refs{\Khrip} is probably the omission of the analogous
terms in the analysis of the photon-graviton scattering vertices.

\vskip0.7cm

\newsec{High Frequency Photon Propagation}

The result of the previous section is the modified light cone formula 
for Ricci flat spacetimes:
\eqn\sectfa{
k^2 ~+~{2\over m^2}~G\Bigl({2k\cdot D\over m^2}\Bigr)~
C_{\m\n\l\r} k^\m k^\l a^\n a^\r ~~=~~0
}
where $G(x)$ is the known function:
\eqn\sectfaa{
G(x) ~~=~~-{1\over2}{\a\over\pi}~\int_0^\infty {ds\over s}~e^{-is}~(is)~
g(s x)
}
where
\eqn\sectfaaa{
g(x) ~~=~~4g_3(0,x,0) ~+~ x~g_9(0,x,0)
}
In terms of the functions $f(x)$ and $F(x_1,x_2,x_3)$ defined in the last 
section, we can write the explicit form for $g(x)$ as:
\eqn\sectfaaaa{
g(x) ~~=~~-{16\over x^2}F(0,x,0) + {3\over x^2}f(x) + \bigl({1\over2} + 
{1\over x}\bigr)f'(x) + {5\over x^2} + {1\over 15}
}
This is plotted below. Analytically, we can show $g(0)= {2\over 90}$, ensuring 
agreement with the Drummond-Hathrell coefficient $c$, while numerically
we find that $g(x)$ approaches an asymptotic value of $0.067$ as $x\rta\infty$.

\vskip0.6cm
\centerline{
{\epsfxsize=7cm\epsfbox{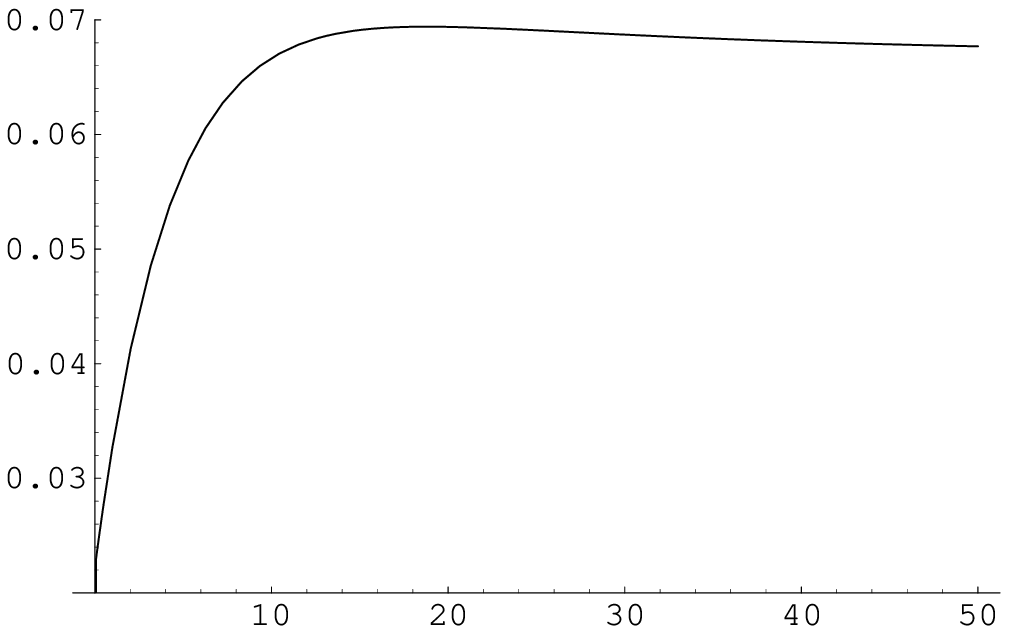}}}
\noindent{\eightpoint Fig.6~~Plot of the function $g(x)$ which enters the formula
for the modified light cone.}
\vskip0.3cm

With this explicit knowledge of the form factor, in principle we have control 
over the high frequency limit of the light cone and the phase velocity.
However, the next difficulty is that in eq.\sectfa, $G$ is a function
of the {\it operator} $k\cdot D$ acting on the Weyl tensor. If $C_{\m\n\l\r}$
is an eigenfunction then $G$ will reduce to a simple function of the eigenvalue
and then Fig.~6 determines its asymptotic behaviour. However, in general this
not true and consequently the interpretation of \sectfa ~is far from
obvious.

At this point, the problem is reduced to differential geometry. The
encouraging feature of \sectfa ~is that the operator $k\cdot D$ simply
describes the variation along a null geodesic. (Notice that because the 
second term is already $O(\a)$, we can freely use the usual Maxwell
relations for the quantities occurring there, e.g.~the results 
$k\cdot D k^\n = 0$ and $k\cdot D a^\n = 0$ derived in section 3.)
The question then becomes what is known about the derivative of the Weyl
tensor along a null geodesic.

As in our previous work, it is convenient to re-cast the light cone 
condition in Newman-Penrose formalism (see, e.g.~refs.{\Inverno,\Ch}
for reviews). This involves introducing a null tetrad $e_a{}^\m$ based on 
a set of complex null vectors $\bigl(\ell^\m, n^\m, m^\m, \bar m^\m\bigr)$.
The components of the Weyl tensor in this basis are denoted by
five complex scalars $\Psi_0, \ldots, \Psi_4$. For example,
$\Psi_0 = -C_{abcd}\ell^a m^b \ell^c m^d$ and 
$\Psi_4 = -C_{abcd}n^a \bar m^b n^c \bar m^d$.

Consider propagation along the null direction $\ell^\m$, i.e.~choose
$k^\m = \ell^\m$. The two spacelike polarisation vectors $a^\m$ and $b^\m$
are related to the $m^\m$ and $\bar m^\m$ null vectors by
$m^\m = {1\over\sqrt2}\bigl(a^\m + i b^\m\bigr)$,
$\bar m^\m = {1\over\sqrt2}\bigl(a^\m - i b^\m\bigr)$.
In this case, the light cone condition can be simply written as
\eqn\sectfb{
k^2 ~+~ {\w^2\over m^2}~ G\Bigl({2\omega\over m^2} \ell^\m D_\m\Bigr)
\Bigl(\Psi_0 + \Psi_0^*\Bigr)~~=~~0
}

We have searched the relativity literature for theorems
on the behaviour of $\ell^\m D_\m \Psi_0$ without finding any
results of general validity, although it seems plausible to us that 
some general properties may exist.
(Notice, for example, that naturally enough this is {\it not} one of the 
combinations that are constrained by the Bianchi identities in Newman-Penrose 
form, as displayed for example in ref.\refs{\Ch}, ch.1, eq.(321).) 
To try to build some intuition,
we have therefore looked at particular cases. The most interesting is
the example of photon propagation in the Bondi-Sachs metric 
\refs{\Bondi,\Sachs} which we recently studied in detail \refs{\Sfour}.
  
The Bondi-Sachs metric describes the gravitational radiation from an
isolated source. The metric is
\eqn\sectfc{
ds^2 = -W du^2 - 2 e^{2\b} du dr + r^2 h_{ij}(dx^i - U^i du)
(dx^j - U^j du)
}
where
\eqn\sectfd{
h_{ij}dx^i dx^j = {1\over2}(e^{2\c} + e^{2\d}) d\theta^2
+ 2 \sinh(\c - \d) \sin\theta d\theta d\phi
+ {1\over2}(e^{-2\c} + e^{-2\d}) \sin^2\theta d\phi^2
}
The metric is valid in the vicinity of future null infinity ${\cal I}^+$.
The family of hypersurfaces $u = const$ are null, i.e. $g^{\m\n}
\pl_\m u \pl_\n u = 0$. Their normal vector $\ell_\m$ satisfies
\eqn\sectfe{
\ell_\m = \pl_\m u ~~~~~~~~~~~~~\Rightarrow~~~~~
\ell^2 = 0, ~~~~~~~~\ell^\m D_\mu \ell^\n = 0
}
The curves with tangent vector $\ell^\m$ are therefore
null geodesics; the coordinate $r$ is a radial parameter along these rays  
and is identified as the luminosity distance. 

The six independent functions  $W,\b,\c,\d,U^i$
characterising the metric have expansions in 
${1\over r}$ in the asymptotic region near ${\cal I}^+$, the coefficients of
which describe the various features of the gravitational radiation.
(See \refs{\Sfour} for a brief summary.)

In the low frequency limit, the light cone is given by the simple
formula \sectab. The velocity shift is quite different for the case of outgoing
and incoming photons \refs{\Sfour}. For outgoing photons, $k^\m = \ell^\m$,
and the light cone is
\eqn\sectff{
k^2 ~~=~~\pm~{4c\w^2\over m^2}~\Bigl(\Psi_0 + \Psi_0^*\Bigr) ~~\sim ~~
O\Bigl({1\over r^5}\Bigr)
}
while for incoming photons, $k^\m = n^\m$,
\eqn\sectfg{
k^2 ~~=~~\pm~{4c\w^2\over m^2}~\Bigl(\Psi_4 + \Psi_4^*\Bigr) ~~\sim ~~
O\Bigl({1\over r}\Bigr)
}

Now, it is a special feature of the Bondi-Sachs spacetime 
that the absolute derivatives of each of the Weyl
scalars $\Psi_0, \ldots, \Psi_4$ along the ray direction $\ell^\m$
vanishes, i.e.~$\Psi_0, \ldots, \Psi_4$ are parallel transported along 
the rays \refs{\Sachs,\Inverno}. In this case, therefore, we have in 
particular:
\eqn\sectfh{
\ell\cdot D ~\Psi_0 ~=~0 ~~~~~~~~~~~~~~~~~~~~~~~~~~~~
\ell\cdot D ~\Psi_4 ~=~0
}
but there is no equivalent simple result for either $n\cdot D ~\Psi_4$
or $n\cdot D ~\Psi_0$.

These results can be easily checked for the simpler related example
of a weak-field plane gravitational wave \refs{\DH}. In this case,
the first identity is trivial since $\Psi_0 = 0$, but in particular  
we can confirm that  $n\cdot D ~\Psi_4 ~\ne~0$ and $\Psi_4$ is not an
eigenfunction. The important result $\ell\cdot D ~\Psi_0 ~=~0$ therefore
appears to be a very special property of the Bondi-Sachs metric
and not an example of a general theorem on derivatives of the Weyl tensor.

Although it seems to be a special case, \sectfh ~is nevertheless important
and leads to a remarkable conclusion. The full light cone condition 
\sectfb ~applied to outgoing photons in the Bondi-Sachs spacetime now 
reduces to
\eqn\sectfi{
k^2 ~+~ {\w^2\over m^2}~ G\bigl(0\bigr)
\Bigl(\Psi_0 + \Psi_0^*\Bigr) ~~=~~ 0
}
since $\ell\cdot D \Psi_0 = 0$. In other words, the low frequency 
Drummond-Hathrell prediction of a superluminal phase velocity $\vp(0)$
is {\it exact} for all frequencies. There is no dispersion, and the 
wavefront velocity $\vp(\infty)$ is indeed superluminal. 

This is potentially a very strong result. According to the analysis
presented here, we have found at least one example in which the
wavefront truly propagates with superluminal velocity. Quantum effects
have indeed shifted the light cone into the geometric spacelike region,
with all the implications that brings for causality.

\vskip0.7cm

\newsec{Propagation in Background Magnetic Fields}

Before accepting the results of the last section as definitive, however,
it is instructive to compare what we have done in this paper for the case 
of a background gravitational field to previously known results on the
propagation of light in a background magnetic field.

The refractive index for photons moving transverse to a homogeneous
magnetic field B has been calculated explicitly as a function of frequency
in two papers by Tsai and Erber \refs{\TEone,\TEtwo}.
They derive an effective action
\eqn\sectga{
\C = -\int dx \biggl[{1\over4}F_{\m\n} F^{\m\n} ~~
+{1\over2} A^\m M_{\m\n}A^\n \biggr]
}
where $M_{\m\n}$ is a differential operator acting on the electromagnetic
field $A^\n$. Writing the corresponding equation of motion and making
the geometric optics ansatz described in section 3, we find the
light cone condition
\eqn\sectgb{
k^2 -  M_{\m\n}(k) a^\m a^\n ~~=~~0
}
In terms of the refractive index $n(\w) = c{|{\ul k}|\over\w} = 
c\vp(\w)^{-1}$, this implies
\eqn\sectgc{
n(\w) = 1 - {1\over2\w^2}  M_{\m\n}(k) a^\m a^\n
}

Denoting $M_{\m\n}a^\m a^\n$ by $M_\parallel, M_\perp$ for the two
polarisations, the complete expression for the birefringent refractive
index is\foot{The extra power of $(is)$ in the prefactor relative to 
eq.\sectfb ~arises because the lowest order terms in the effective action
contributing to \sectga ~are of 4th order in the background fields, 
i.e. $O(F^4)$, compared
to the 3rd order terms of $O(RFF)$ in the gravitational case.
For background magnetic fields, the analogue of the Drummond-Hathrell
low frequency action is the familiar Euler-Heisenberg action (see, 
e.g.~ref.\refs{\Sone}).}:
\eqnn\sectgd
$$\eqalignno{
n_{\parallel,\perp}(\w) ~~&=~~ 1 ~-~ {1\over2\w^2}~M_{\parallel,\perp} \cr
&=~~1~-~ 
{\a\over4\pi} \Bigl({eB\over m^2}\Bigr)^2 ~
\int_{-1}^1 du \int_0^\infty ds ~s ~
N_{\parallel,\perp}(u,z)~ e^{-is\bigl(1 + s^2 \W^2 P(u,z)\bigr)} \cr
{}&{}& \sectgd \cr}
$$
where $z = {eB\over m^2}s$ and $\W = {eB\over m^2}{\w\over m}$. 
The functions $N$ and $P$ are given by:
\eqn\sectfe{
P(u,z) = {1\over z^2} \Bigl({\cos zu - \cos z\over 2z\sin z} 
- {1-u^2\over4}\Bigr)
~~=~~{1\over12}(1-u^2) + O(z^2)
}
and
\eqnn\sectgf{
$$\eqalignno{
N_{\parallel} &= -{\cot z\over2z}\Bigl(1-u^2 + {u\sin zu\over\sin z}\Bigr)
+{\cos zu\over 2z\sin z}~~
=~~ {1\over4} (1-u^2)(1-{1\over3}u^2) + O(z^2) \cr
{}&{}\cr
N_{\perp} &= -{z\cos zu\over 2\sin z} +{zu\cot z \sin zu\over 2\sin z} +
{z(\cos zu -\cos z)\over\sin^2 z}~~
=~~ {1\over8} (1-u^2)(1+{1\over3}u^2) + O(z^2) \cr
{}&{}& \sectff \cr}
$$

In the weak field, low frequency limit, we can disregard the function $P$
and consider only the lowest term in the expansion of $N$ in powers of $z$.
This reproduces the well-known results (see also \refs{\Adler,\Sone}):
\eqn\sectgg{
n_{\parallel,\perp} ~ \sim_{\w\rta 0}~ 
1 + {\a\over4\pi} \Bigl({eB\over m}\Bigr)^2~
\Bigl[{14\over45}_\parallel, {8\over45}_\perp \Bigr]
}
The weak field, high frequency limit is analysed in ref.\refs{\TEtwo}.
It is shown that 
\eqn\sectgh{
n_{\parallel,\perp} ~ \sim_{\w\rta \infty}~
1 - {\a\over4\pi} \Bigl({eB\over m}\Bigr)^2~
\bigl[{c}_\parallel, {c}_\perp \bigr]~ \W^{-{4\over3}}
}
where the numerical constants are $\bigl[{c}_\parallel, {c}_\perp \bigr]
= {3^4 3^{1\over3}\over7}\sqrt{\pi} \C({2\over3})^2 \bigl(\C({1\over6})
\bigr)^{-1}~[3_\parallel,2_\perp]$.

The complete function $n(\w)$ is sketched in Fig.~7. It shows exactly the
features found in the simple absorption model described in section 2.

\vskip0.2cm
\centerline{
{\epsfxsize=6cm\epsfbox{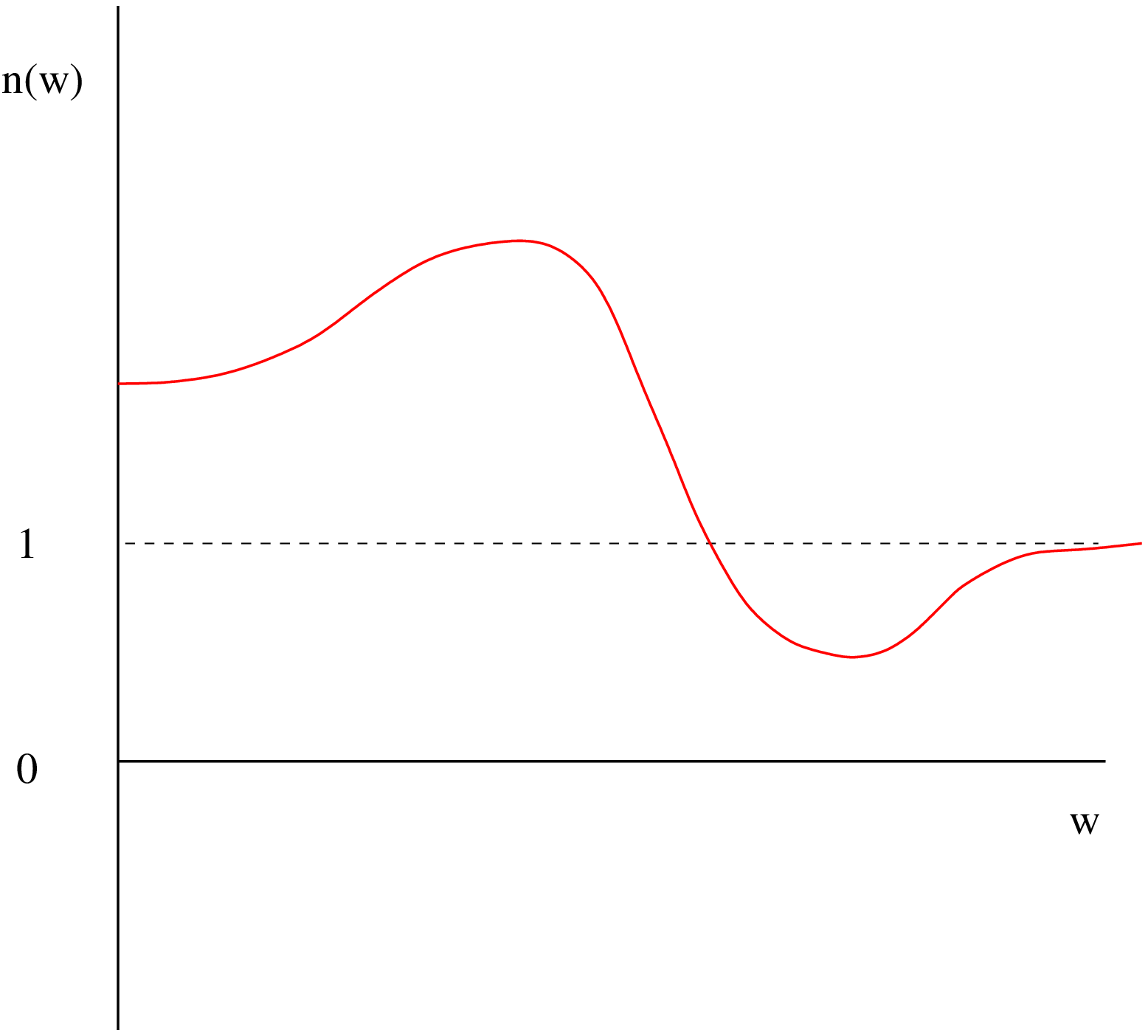}}}
\noindent{\eightpoint Fig.7~~Sketch of the frequency dependence of the 
refractive index $n(\w)$ for light propagating in a background magnetic field. 
The crossover point is at $\W \sim 1$.} 
\vskip0.2cm

\noindent In particular, the phase velocity $\vp(\w)$ begins less than 1
at low frequencies, showing birefringence but conventional subluminal
behaviour. In the high frequency limit, however, the phase velocity
approaches $c$ from the superluminal side with a $\w^{-{4\over3}}$ 
behaviour. 

All this is by now standard and in line with our expectations. What we are 
interested in here, however, is the comparison between this calculation of 
$\vp(\w)$ via the effective action \sectga ~and the gravitational calculation 
of sections 5 and 6, in particular the origin of the non-analytic high 
frequency behaviour \sectgh.

The important observation is that the exponential term involving $\W^2 P$
is crucial in obtaining the high frequency limit. However, if we made a 
literal expansion of the effective action in powers of the magnetic field
$(eB/ m^2)$, this term would be regarded as higher order and discarded. 
It {\it is} important however, because it involves the product of the
field and the frequency, $(eB\w/m^3)$, and this is {\it not} small in
the interesting high frequency region. The exponent in the effective action
is vital in describing the high frequency propagation.

Now compare with our construction of the effective action in a background
gravitational field. The action \sectdg ~is obtained by expanding in the 
curvature and keeping only terms of $O(R)$. This appears to be analogous
to the expansion of $N$ in \sectgd. We have found a non-trivial extension of
the zero-frequency Drummond-Hathrell result by taking into account derivatives
of the curvature. These are essential in the gravitational case because there 
is no change in the light cone for a constant curvature metric 
(since this is totally isotropic), whereas there is an interesting
birefringent effect even for a constant magnetic field. By restricting
the effective action rigorously to terms of first order in the curvature,
however, we seem to have missed the analogue of the exponent terms in \sectgd,
which would be characterised by the not necessarily small parameter 
$(R\w/ m^3)$.  

The conclusion of this comparison is therefore discouraging. It seems that
despite its complexity, the effective action \sectdg ~may still not be 
sufficiently general to encode the high frequency behaviour of photon 
propagation.

\vskip0.7cm

\newsec{Conclusions}

In this final section, we attempt to synthesise what has been learnt from
this investigation and identify what further work is required for a
complete resolution of the dispersion problem for the propagation of 
photons in gravitational fields.

In view of our experience with the background magnetic field problem,
it is natural to assume that the full 
effective action governing propagation in a background gravitational 
field takes an analogous form and the modified light cone can be
written heuristically as
\eqn\sectha{
k^2 + {\a\over\pi} \int_0^\infty {ds\over s}~(is)~{\cal N}(s,R)~e^{-is\bigl(
1 + s^2\W^2 {\cal P}(s,R)\bigr)}~~=~~0
}
where both ${\cal N}$ and ${\cal P}$ can be expanded in powers of curvature,
and derivatives of curvature, with appropriate powers of $s$. The frequency
dependent factor $\W$ would be $\W \sim {R\over m^2}{\w\over m^2} \sim
O({\l_c^3\over\l L^2})$, where `$R$' denotes some generic 
curvature component. If this is true, then an expansion of the effective
action to first order in $O({R\over m^2})$ would not be sensitive to the 
${\cal P}$ term in the exponent. The Drummond-Hathrell action would correspond 
to the leading order term in the expansion of ${\cal N}(s,R)$ in powers of 
${R\over m^2}$ neglecting derivatives, while our improved effective action
sums up all orders in derivatives while retaining the restriction to
leading order in curvature.  

The omission of the ${\cal P}$ term would be justified only in the 
limit of small $\W$, i.e.~for ${\l_c^3\over\l L^2} \ll 1$. Neglecting this 
therefore prevents us from accessing the genuinely high frequency limit
$\l \rta 0$ needed to find the asymptotic limit $\vp(\infty)$ of the phase 
velocity. If eq.\sectha ~is indeed on the right lines, it also looks inevitable
that for high frequencies (large $\W$) the rapid variation in the exponent
will drive the entire heat kernel integral to zero, restoring the usual 
light cone $k^2=0$ as $\w\rta\infty$. However, this is simply to assert
that in this respect the gravitational problem behaves in the same way as
the magnetic field background, and while this seems plausible in relation to
the $\w\rta\infty$ limit we should be cautious:~systematic cancellations or
special identities (c.f.~section 6) could change the picture, and we should
remember that it was not foreseen that the low frequency behaviour of $\vp(0)$ 
would differ so radically from other background field calculations as to 
produce superluminal velocities. So although the available evidence seems to
point strongly towards a restoration of the usual light cone in the high 
frequency limit, this inference should be made with some caution.

Unfortunately, the quantum field theoretic calculation required to settle
the issue by evaluating the `${\cal P}$' type correction to the exponent
in \sectha ~looks difficult in general, at least comparable to the evaluation 
of the effective action `${\cal N}$' terms in section 4. However, if we are
only interested in the $\w\rta\infty$ limit, it may be sufficient just to 
perform some leading order type of calculation to establish the essential
$\bigl(1 + s^2 \W^2 {\cal P}\bigr)$ structure of the correction. Another possible 
simplification, which would be interesting in its own right, would be to 
reformulate the effective action calculation from the outset in Newman-Penrose
form, thereby removing the plethora of indices arising in calculations 
involving the curvature tensors. Simple cases, such as black holes, in which
only one Weyl scalar is non-vanishing might prove particularly tractable. 

However, even if this picture is correct and the light cone is eventually
driven back to $k^2=0$ in the high frequency limit, the analysis in this
paper still represents an important extension of the domain of validity of
the superluminal velocity prediction of Drummond and Hathrell. Recall
from section 5 that the constraint on the frequency for which the 
superluminal effect is in principle observable is ${\l_c^2\over\l L} \gg 1$.
Obviously this was not satisfied by the original $\w\sim 0$ derivation.
However, our extension based on the `all orders in derivatives' effective
action does satisfy this constraint. Combining with the restriction
${\l_c^3\over \l L^2} \ll 1$ in which the neglect of the ${\cal P}$ type
corrections is justified, we see that there is a frequency range
\eqn\secthb{
{\l_c\over L} ~~\gg ~~ {\l\over \l_c} ~~\gg~~ {\l_c^2\over L^2}
}
for which our expression \sectfa
\eqn\sectfa{
k^2 ~+~{2\over m^2}~G\Bigl({2k\cdot D\over m^2}\Bigr)~
C_{\m\n\l\r} k^\m k^\l a^\n a^\r ~~=~~0
}
for the modified light cone is valid and predicts observable effects.

Since this formula allows superluminal corrections to the light cone,
we conclude that superluminal propagation has indeed been established as
an observable phenomenon even if, as seems likely, causality turns out to 
respected through the restoration of the standard light cone $k^2=0$ 
in the asymptotic high frequency limit.

\vskip1cm

\newsec{Acknowledgements}

I would like to thank I. Drummond, H. Gies and W. Perkins for many helpful 
conversations on this topic. I am especially grateful to A. Dolgov, V. Khoze 
and I. Khriplovich for their help in obtaining and interpreting 
ref.\refs{\Leon}. This research is supported in part by PPARC grant 
PPA/G/O/2000/00448.

\vfill\eject

\listrefs

\bye